\documentstyle[12pt,epsfig]{article}
\begin{document}
\newcommand \be  {\begin{equation}}
\newcommand \bea {\begin{eqnarray} \nonumber }
\newcommand \ee  {\end{equation}}
\newcommand \eea {\end{eqnarray}}


\def\gmin{\mathcal{G}}
\def\tmin{\mathcal{T}}
\def\vmaxi{V_{1}}
\def\smaxi{\mathcal{S}_{1}}

\title{Statistical mechanics of a single particle in
a multiscale random potential: \\ Parisi landscapes
in finite dimensional Euclidean spaces}

\vskip 0.2cm

\author{Yan V Fyodorov$^{1}$ and Jean-Philippe Bouchaud$^{2}$}
\maketitle

\noindent\small{ $^1$ School of Mathematical Sciences, University
of Nottingham, Nottingham NG72RD, England\\$^2$ Science \&
Finance, Capital Fund Management 6-8 Bd Haussmann, 75009 Paris,
France.
}

\begin{abstract}
{\small We construct a $N-$dimensional Gaussian landscape with
multiscale, translation invariant, logarithmic correlations and
investigate the statistical mechanics of a single particle in this
environment.  In the limit of high dimension $N\to \infty$ the
free energy of the system and overlap function are calculated
exactly using the replica trick and Parisi's hierarchical ansatz.
In the thermodynamic limit, we recover the most general version of
the Derrida's Generalized Random Energy Model (GREM). The
low-temperature behaviour depends essentially on the spectrum of
length scales involved in the construction of  the landscape. If
the latter consists of $K$ discrete values, the system is
characterized by a K-step Replica Symmetry Breaking solution. We
argue that our construction is in fact valid in any finite spatial
dimensions $N\ge 1$.  We discuss the implications of our results
for the singularity spectrum describing multifractality of the
associated Boltzmann-Gibbs measure. Finally we discuss several
generalisations and open problems, such as the dynamics in such a
landscape and the construction of a Generalized Multifractal
Random Walk.}
\end{abstract}

\vskip 0.1cm

PACS numbers 64.60.De, 64.60.Al \\

\section{Introduction}

Ever since the seminal paper of Goldstein in 1969
\cite{Goldstein}, the idea of energy landscapes pervades the
theoretical description of glasses, disordered systems, proteins,
etc., see  \cite{Wales,KL,Cavagna,FW} and references therein. The
general idea is to describe the statics and dynamics of the whole
system, or one of its subparts, by a single point particle moving
in a random potential, which encodes the complexity of the
original system. The hope then is to be able to classify the
possible classes of random potential and to establish generic,
universal properties, in the spirit of Random Matrix Theory. In
this respect, the Parisi solution for spin-glasses is fascinating:
it reveals that in this case the energy landscape has a
surprisingly complex, hierarchical structure of valleys within
valleys within valleys, etc. \cite{Parisi}. It is often argued
that this construction is very specific not only to infinite range
spin models, but also to infinite dimensional landscape models. In
particular, the ultrametric properties of Parisi landscapes seem
at first sight hardly compatible with a finite dimensional,
translation invariant random function.\footnote{In fact, random
potentials with a hierarchical Parisi structure can be constructed
in finite dimensional space in a kind of ad-hoc way, by following
step by step the real-space interpretation Replica Symmetry
Breaking: see \cite{BBM} for a discussion of this point.} In this
paper, we provide an explicit construction of a Gaussian random
potential in Euclidean, $N$ dimensional spaces, with a specific
form of long-ranged correlations which reproduces all the features
of Parisi landscapes. More precisely, we show that the
thermodynamics of a single particle in a multiscale,
logarithmically correlated potential is exactly described by
Derrida's Generalized Random Energy Model (GREM, \cite{GREM}),
with an arbitrary (possibly infinite) number of levels of
hierarchy. Although our proof concerns, strictly speaking, the
limit $N \to \infty$, we are confident that our results hold in
arbitrary finite dimension $N \geq 1$. This conviction is built
both on physical arguments and on the beautiful results of
Carpentier and Le Doussal \cite{CLD} on the monoscale version of
our model in finite dimensions, which, as shown recently, match
precisely the exact results of the same model when $N \to \infty$
\cite{FS}.

As is well-known, the Gibbs-Boltzmann measures in systems with
disorder often possess the interesting property of being
multifractal, see the papers \cite{CLD,2d,MG}. In fact, this
property is not unrelated to the multifractality of the
wavefunctions in disordered electronic systems (see \cite{ME} for
a comprehensive discussion of the last topic and further
references). The investigation of multifractal measures of diverse
origin has been a very active field of research in various
branches of physics for about two decades now\cite{PV,BMD,ME}.
From this point, we show that our results imply, in particular,
the possibility of a rather rich and unusual behaviour of the
singularity spectrum describing multifractality of the
Gibbs-Boltzmann measure arising in our multiscale logarithmic
model.

Another closely related aspect is that the monoscale model of
Carpentier and Le Doussal is known to be the building block in the
construction of an exact {\it multifractal random walk} (MRW),
proposed by Bacry, Muzy \& Delour to describe financial time
series \cite{BMD}; the Boltzmann weight in one language
corresponds to the local volatility in the other. So our extended
multiscale model can also be interpreted as the construction of a
Generalized Multifractal Random Walk (GMRW), in the same sense as
the GREM generalizes the Random Energy Model. Physically, the
result is that the $n$-th moment of the distance travelled by the
random walk scales with an exponent that does not only depend on
$n$ but also on the {\it epoch}, ie the logarithm of the time lag.

The outline of the paper is as follows: in Sect.\ref{sect2} we
introduce landscape models in full generality, recall the set of
previously established results and discuss how they can be
understood qualitatively. We then focus on the case of
logarithmically correlated landscapes, summarize the recent
findings of ref. \cite{FS} and their relation with the results of
Carpentier and Le Doussal \cite{CLD}. In Sect. \ref{sect3}, we
define precisely our multiscale random landscape model, discuss
its physical motivation and detail our analytical calculations in
the large dimensional $N \to \infty$ limit, where we recover
exactly the GREM results.  We end this section by describing the
most interesting features of the multifractality spectrum of the
associated Boltzmann-Gibbs measure implied by our results for the
multiscaled landscapes. Finally, in Sect. \ref{sect4} we put
forward several conjectures about the {\it dynamics} of a point
particle in such landscapes, open problems and generalisations,
concerning for example Generalized Multifractal Random Walks. Some
more technical points are relegated to the Appendix.

A short account of some of the results of the present paper was
presented in \cite{FBshort}.

\section{Thermodynamics of a particle in a random potential}
\label{sect2}
\subsection{General discussion}

As alluded  to in the introduction, the ``toy-model'' of a
classical particle in a random potential exhibits a rich variety
of behaviour which mimics many of the dynamical and
thermodynamical properties of glassy systems. The model is defined
as follows: the position of the particle, confined inside an
$N-$dimensional spherical box of radius $L$, is described by the
coordinate vector ${\bf r}=(r_1,...,r_N),\,\, |{\bf r}|\le L$. It
feels a random potential  $V({\bf r})$, which we conventionally
choose to be Gaussian-distributed with zero mean, and with
covariance chosen to be isotropic, translation invariant and with
a well-defined large $N-$limit:
\begin{equation}\label{potential}
\left\langle V\left({\bf r}_1\right) \, V\left({\bf
r}_2\right)\right\rangle_V=N\,f\left(\frac{1}{2N}({\bf r}_1-{\bf r}_2)^2\right)\,.
\end{equation}
In Eq.(\ref{potential}) and henceforth the notation
$\left\langle\ldots\right\rangle_V$  stands for an ensemble
average over the random potential, and $f(u)$ is a function of
order unity belonging to the so-called class ${\cal D}_{\infty}$
described in detail, e.g., in the book by Yaglom \cite{Yaglom}.
The functions $f(u)\in {\cal D}_{\infty}$ are such that they
 represent covariances of an isotropic random field
 for {\it any} spatial dimension $N\ge 1$. There are two essentially different
  types of such functions. The first type corresponds
to genuine isotropic random fields, and those $f(u)$ are
characterized by a non-negative normalizable ``spectral density
function'' $\tilde{f}(k)\ge 0, \, k\ge 0$ in terms of which $f(u)$
is represented as (see \cite{Yaglom} p.354):
\begin{equation}\label{shortranged}
f(u)=\int_0^{\infty}e^{-k^2 u} \tilde{f}(k) dk, \qquad
f(0)=\int_0^{\infty}\tilde{f}(k)dk<\infty\,.
\end{equation}
In particular, $f(u)$ is decreasing and convex, i.e. satisfies
$f'(u)<0, f''(u)>0$ $\forall u\ge 0$ , and in addition
$f'(u\to\infty)=0$. Here and below the number of dashes indicates
the order of derivatives taken. A few important families of such
functions listed in \cite{Yaglom} are, e.g., (i)
$f(u)=Ce^{-au^{\gamma}}$, $0\le \gamma \le 1,\, C>0,\, a>0$ (ii)
$f(u)=C/(a+u)^{\gamma}$, $\gamma>0,\, C>0,\, a>0$ and (iii)
$f(u)=C(au)^{\gamma/2}K_{\gamma}\left(\sqrt{au}\right)$,
$\gamma>0,\, C>0,\, a>0$, where $K_{\gamma}(x)$ stands for the
modified Bessel (a.k.a. Macdonald)  function. In the physical
literature the random fields of that type are frequently called
potentials with short-ranged (SR) correlations.

 The second type of covariances occurs in the situation when the normalization integral
$\int_0^{\infty}\tilde{f}(k)dk$ diverges. It corresponds to
long-ranged (LR) random fields with isotropic {\it increments}
also known as {\it locally} isotropic random fields (see e.g.
\cite{Yaglom}, p.438). The spectral density function now
 must satisfy the condition $\int_0^{\infty}\frac{k^2}{k^2+1}\tilde{f}(k)dk<\infty$
 which allows one to prove that in any dimension $N\ge 1$ there exists a random field
whose {\it structure function}
$\frac{1}{2}\langle\left(V(0)-V({\bf r})\right)^2
\rangle_V=f(0)-f(u)$ is given by
\begin{equation}\label{longranged}
f(0)-f(u)=\int_0^{\infty} dk (1-e^{-k^2u})\tilde{f}(k)dk+Au\,,
\quad A\ge 0\,.
\end{equation}
In what follows we will impose an additional requirement $f'(u\to
\infty)=0$, which ensures $A=0$ (no external driving force acting on
the particle). It is also easy to convince oneself that in the
present model the difference between the covariance and the
structure function, i.e. the value of $f(0)$, is immaterial for the
free-energy calculations. The most widely-known example of the
locally isotropic LR field is the so-called {\it self-similar}
random field, see \cite{Yaglom} p. 441, characterized by the
spectral density $\tilde{f}(k>0)=k^{-2\gamma-1},\, 0<\gamma<1$. The
corresponding covariance behaves as
\begin{equation}\label{longranged1}
f(u)=f(0)-C_{\gamma}u^{\gamma}\,.
\end{equation} In particular, for $N=1$ and $\gamma=1/2$ this is the example of a
simple Brownian motion for the potential, corresponding the celebrated Sinai model.

After specifying in detail the class of random potentials involved
in our construction, let us turn to thermodynamics of the model
characterized by the following partition function and the
corresponding free-energy:
\begin{equation}\label{freeendef}
F=-T\,\langle \ln{Z}\rangle_V ,\, \quad Z=\int_{|{\bf r}|\le L}
\exp{-\beta V({\bf r})}\, d {\bf r}\,,
\end{equation}
where $\beta=\frac{1}{T}$ stands for the inverse temperature. A
variant of the model consists in replacing the spherical box
$|{\bf r}|\le L$ by a confining harmonic potential $-\mu {\bf
r}^2/2$. Such a model has been studied extensively since the
mid-eighties. It was originally proposed in $N=1$ dimension as a
toy-model for a randomly pinned domain wall \cite{Villain} or of a
directed polymer in a random potential \cite{BO} and studied using
a variety of methods \cite{MC,BK,CLD,MLD},  some of them being
exact. Another case where analytical calculations can be performed
is the high-dimensional limit $N \to \infty$ \cite{MP,Engel},
where a Gaussian variational ansatz with Replica Symmetry Breaking
becomes exact. One finds that the nature of the low temperature
phase is essentially dependent on the behaviour of the covariance
at large distances \cite{MP,Engel}. Namely, for typical
short-ranged correlated potential the description of the low
temperature phase was found to require the so-called one-step
replica symmetry breaking (1RSB) scheme of Parisi. In contrast,
for the potentials growing as $u^{\gamma}$, see
Eq.(\ref{longranged1}) , the full infinite-hierarchy replica
symmetry breaking (FRSB) scheme has to be used. In fact, this
problem was reconsidered recently in \cite{FS} in much detail
using an alternative method that directly focusses on the degrees
of freedom relevant in the limit $N\to \infty$, and employs the
Laplace (aka saddle-point) method for evaluating the integrals. In
the limit $N \to \infty$, one actually finds a true phase
transition as a function of temperature provided the radius of the
confining sphere $L$ is scaled as $R \sqrt{N}$. The effective size
$R<\infty$ (which is accidentally just half of the length of an
edge of the cube inscribed in this sphere) is then used as the
main control parameter of the model. The chosen scaling $L\sim
\sqrt{N}$ formally stems from the property of the argument of the
correlation function Eq.(\ref{potential}) to become of order of
unity for separations of order of $\sqrt{N}$. More importantly, it
simultaneously ensures that the volume
$V_L=\pi^{N/2}\frac{L^N}{\Gamma(N/2+1)}$ of our spherical sample
retains in the limit $N\gg 1$ the natural scaling with size $R$
and dimension $N$: $\ln{V_L}=N\ln{(R/R_0)}+$ smaller terms, with
$R_0$ being a constant of order of unity. Such a behaviour is
essential since the phase transition is physically induced by a
competition between entropic effects, which tend to delocalize the
particle over the sphere, and the minima of the random potential
which tend to attract, and possibly to localize the particle over
a {\it finite} number of favorable sites. In the short-range case,
the number of effectively independent "sites" is of the order of
$V_L\propto R^N$ which ensures, thanks to the $N$ factor in front
of $f(u)$ in Eq. (\ref{potential}), that the minimum of the
Gaussian potential scales as $\sqrt{N} \sqrt{\ln R^N} \propto N$.
This indeed can compete with the entropy of the order $N \ln R$.
All these arguments demonstrate that indeed $R$ is the most
natural measure of the sample size.

The fact that a true thermodynamic transition exists for a finite
sample size $R<\infty$ is obviously a somewhat pathological
feature of the limit of infinite dimension  $N \to \infty$ taken
first. Indeed, in that limit the total number of thermodynamic
degrees of freedom is infinite even for finite $R$. At the same
time, at any finite spatial dimension $N<\infty$ phase transitions
only may occur in the thermodynamic limit of infinite sample size $R \to \infty$.
From this point of view it is natural to inspect the $R\to \infty$
behaviour of the transition temperature $T_c(R)$ at which the
system experiences a continuous breakdown of replica symmetric
solution. For the present model it reads, according to \cite{FS}:
\be\label{Tc} T_{c}(R)\approx R^2\sqrt{f''(R^2)} \ee in
agreement with a similar result for the confining quadratic
potential case \cite{MP,Engel}. Therefore, one finds that $T_c(R)$
tends for $R \to \infty$ either to zero in the SR
case\footnote{The situation is slightly more complicated, as for
SR case there exists another temperature $T_1(R)>T_c(R)$ where the
system experiences a discontinuous breakdown of replica symmetry.
However one can show that $T_1(R\to \infty)\to 0$ as well
\cite{FS}.}, or to infinity in the LR case. A more detailed
analysis shows correspondingly that the large $R$ behaviour of the
free-energy is $F(T)|_{R\to \infty}\sim -T\ln{R^N}$ in the SR case
where entropy dominates, and $F(T)|_{R\to \infty}\sim
-NR^{\gamma}$ in the LR case where the deepest minimum dominates.

\subsection{Logarithmically correlated potentials}
\label{sect2end} From Eq. (\ref{Tc}) above, $T_c(R)|_{R\to
\infty}$ appears to have a well defined limit when $f''(u) \sim
u^{-2}$ for large $u$, corresponding to a logarithmically growing
correlation function of random potential, a case overlooked in
previous studies \cite{MP,Engel}. The peculiarities of that case
can be traced in a few different ways. To this end it is
appropriate to mention a precise mathematical criterion proposed
recently in \cite{FS} to classify statistical mechanics behaviour
induced by SR vs LR correlated potentials. The criterion uses the
notion of the so-called Schwarzian derivative
$\{f'(u),u\}=-S(u)/[f''(u)]^2$, where $S(u)$ is expressed in terms
of $f(u)$ as
\begin{equation}\label{2}
S(u)=\frac{3}{2}\left[f'''(u)\right]^2-f''(u)f''''(u)\,.
\end{equation}
In terms of $S(u)$ it was demonstrated that
\begin{itemize}
\item any potential whose covariance function satisfies the condition\\
$S(u)>0\,\, \forall u\ge 0$ must have a 1RSB low temperature phase.
It is easy to check that such a situation includes, in particular,
the standard families of the SR potentials (i) and (ii) listed after
Eq.(\ref{shortranged}). For a general case of SR fields with
covariances defined via Eq.(\ref{shortranged}) one always has
$S(u)>0$ for large enough values of $u$, which is the most essential
range in the thermodynamic limit.

\item Any potential whose correlation function satisfies the
condition\\
$S(u)<0\,\,\forall u\ge 0$ must necessarily have the FRSB low
temperature phase. This condition holds for the standard LR
correlation functions of the type Eq.(\ref{longranged1}), i.e. for
$f(u)=f_0-g^2(u+a^2)^{\gamma}$, with $0<\gamma<1$ and
$f_0-g^2\,a^{2\gamma}>0$. It is natural to conjecture that typical
LR random fields with independent increments should be of this type
for large enough $u$.
\end{itemize}

Clearly, the above criterion naturally singles out as a special
marginal case random potentials satisfying $S(u)=0$. The only
function satisfying this condition globally, i.e for all $u\ge 0$,
and satisfying also the requirement $f'(u\to \infty)=0$ is indeed
given by a {\it logarithmic} correlation function, of the type
considered by Carpentier and Le Doussal in finite dimensions
\cite{CLD}:
\begin{equation}\label{2c}
f(u)=f_0-g^2\ln{(u+a^2)}\,,
\end{equation}
where $g$ and $a$ are given constants, and $f_0$ is such that
$f(0)=f_0-g^2\ln{(a^2)}>0$\footnote{As noted above, the value of
$f(0)$ is irrelevant for the thermodynamics of the system, so that
$f_0$ can be dropped from the calculations. We systematically
disregard such constants in the rest of the paper.}. Let us stress
that the expression Eq.(\ref{2c}) is a legitimate covariance
function belonging to the ${\cal D}_{\infty}$ class of LR locally
isotropic fields, Eq.(\ref{longranged}). Indeed, it corresponds to
the spectral density of the form
$\tilde{f}(k)=\frac{2g^2}{k}e^{-a^2k^2}$, which satisfies the
required condition
$\int_0^{\infty}\frac{k^2}{k^2+1}\tilde{f}(k)dk<\infty$.

In such a case of logarithmically-correlated potentials the solution found in
\cite{FS} has features of both the
SR-1RSB and LR-FRSB regimes. The critical temperature $T_c$ remains finite for large systems,
and is given by: \be T_c(R\to \infty)=g. \ee
Physically, the minima of the potential
 now typically behave as:
\be
V_{\min}(R) \sim -g \sqrt{N} \sqrt{2\ln R^N} \sqrt{\ln(a^2+R^2)} \sim -2 g N \ln R,
\ee
while the entropy contribution is $-TN \ln R$, suggesting that indeed some change of
physics should take place when $T \sim g$ (see \cite{CLD} for further elaboration of this argument).

Furthermore, it is easy to check that the free-energy expression
found in \cite{FS} for arbitrary $R$ and $a$ reduces in the limit
$R\gg a$ to that of the famous Random
Energy Model \cite{REM}:
\begin{equation}\label{REM}
-\frac{1}{N}F(T)|_{R \gg a}=\left\{\begin{array}{l}
T(1+g^2/T^2)\,\ln{R},\quad T>T_c\\ 2g\ln{R},\qquad\qquad\qquad T<T_c
\end{array}\right.
\end{equation}
Interestingly, these results coincides {\it precisely} with the
Renormalisation Group results of Carpentier and Le Doussal in finite dimensions
(up to a rescaling of their coupling constant $g=\sqrt{\sigma}$ by a factor $\sqrt{N}$,
 as indicated by Eq. (\ref{potential})). The interpretation is the same as for the REM:
 below $T_c$, the partition function becomes dominated by a finite number of
 sites where the random potential is particularly low, and where the particle ends up spending most of its time
 \cite{MB,BenArous}.  For a more quantitative description of the particle localization, useful in the following,
it is natural to employ the overlap function defined as the mean
probability for two {\it independent} particles placed in the same
random potential
 to end up at a given distance to each other. Denoting the
 scaled Euclidean distance (squared) between the two points in the sample as
 ${\cal D}$, and employing the Boltzmann-Gibbs equilibrium measure $p_{\beta}({\bf
r})=\frac{1}{Z(\beta)}\exp{-\beta V({\bf r})}$ the above
probability in thermodynamic equilibrium should be given by
 \begin{equation}\label{overlap}
 \pi({\cal D})=
 \left\langle\int_{|{\bf r}_1|<L}\,d{\bf r}_1\,p_{\beta}({\bf
r}_1)\int_{|{\bf r}_2|<L}d{\bf r}_2\,p_{\beta}({\bf r}_2)\,
 \delta\left({\cal D}-\frac{1}{2N}|{\bf r}_1-{\bf
 r}_2|^2\right)\right\rangle_{V}\,
\end{equation}
where here and henceforth $\delta$ denotes the Dirac's
$\delta$-function. The disorder averaging in (\ref{overlap}) can
be calculated following the same standard steps of the replica
approach as the free energy itself. For convenience of the reader
we sketch the procedure for the present model in the Appendix A.

 With the function $\pi({\cal D})$ in hand we can ask, in particular,
  what is the probability for the particle in logarithmically correlated potential
   to end up at ${\cal D}=O(a^2)$, i.e.  at a distance of order of
  the small cutoff scale. The answer turns out to be zero in the high-temperature phase
   $T > T_c$, confirming the particle delocalization
   over the sample. In contrast, in the lower temperature phase $T <
T_c$ the probability is finite: $\pi\left(O(a^2)\right)=1-T/T_c$,
again in full agreement with REM calculation \cite{REM,BM,CB,CLD}.

A logarithmic growth of the variance of the potential might look an academic oddity,
but in fact is not, and appears naturally in various systems of actual physical interest.
We warmly recommend
the paper of Carpentier and Le Doussal \cite{CLD}, which discusses in detail the
connection to many other interesting and important physical problems, like directed polymers in random
environment \cite{DS}, or a quantum particle in a random magnetic field \cite{2d}.
In the next section, we introduce and study a very natural, multiscale generalisation of this model.

\section{A multiscale logarithmic potential}
\label{sect3}
\subsection{Motivation and definition of the model}

The main observation of the present paper is that the above picture, despite looking rather complete,
 still misses a rich class of possible behaviour that survives in the thermodynamic limit
$R \to \infty$. Namely, given any increasing positive function $\Phi(y)$ for $0<y<1$, we
demonstrate below that if one considers potential correlation functions $f(u)$ which take the following
scaling form
\begin{equation}\label{scalingln}
f(u)=-2 \ln{R}\,\, \Phi\left(\frac{\ln{(u+a^2)}}{2\ln{R}}\right),
\quad 0 \le u+a^2 <R^2,
\end{equation}
the thermodynamics of our system in the limit $R\to \infty$ is precisely equivalent to that of
celebrated Derrida's Generalized
Random Energy Model (GREM)\cite{GREM}. The REM-like case discussed above turns out to be
only a (rather marginal) representative of
this class corresponding to specific choice of the scaling function $\Phi(y)=g^2 y$.

Let us explain the motivation of the above form, which will make
the physical interpretation of the results (as well as some
technical calculations) quite transparent. The idea is to write
$V({\bf r})$ as a (possibly infinite) sum of $K$ independent
Gaussian potentials:\be\label{multidef} V({\bf r})=\sum_{i=1}^K
V_i({\bf r}), \ee each with a simple logarithmic covariance as in
(\ref{2c}): \be \left\langle V_i \left({\bf r}_1\right) \, V_j
\left({\bf r}_2\right)\right\rangle_V=\delta_{i,j} N\,f_i
\left(\frac{1}{2N}({\bf r}_1-{\bf r}_2)^2\right)\,, \qquad
f_i(u)=-g_i^2\ln{(u+a^2+a_i^2)}\,, \ee each with its own strength
constant $g_i$ and small-scale cutoff $a_{i}$, which we choose to
grow as a power-law of the system size:\footnote{One could in fact
multiply this power law behaviour by a slow function of $R$ with
no impact on the following results in the limit $\ln R \to
\infty$.} $a_{i}=R^{\nu_i}$ with $0 \leq \nu_i \leq 1$. Taking the
continuum limit $K \to \infty$ with a certain density $\rho(\nu)$
of exponents $\nu_i$, we end up with:
\begin{equation}\label{scalingln1}
f(u)=-\int_{0}^{1}\rho(\nu) g^2(\nu) \ln{\left(u+a^2+R^{2\nu}\right)}\,d\nu,
\quad 0\le x\le R^2.
\end{equation}
Now, introducing $u+a^2 \equiv R^{2y}$ and identifying with Eq.
(\ref{scalingln}) in the $R \to \infty$ limit, we find that the
function $\Phi$ has the following representation: \be \Phi(y) = y
\int_0^y \rho(\nu) g^2(\nu) \,d\nu +  \int_y^1 \nu \rho(\nu)
g^2(\nu) \,d\nu, \ee the previous REM case corresponding to
$\rho(\nu)=\delta(\nu)$. Note also that in this representation,
$\Phi'(y)= \int_0^y \rho(\nu) g^2(\nu) \,d\nu \geq 0$, and also
$\Phi''(y) \geq 0$. The main result of this work is the following:
depending on the nature of the spectrum of the exponents $\nu$,
discrete or continuous, we will recover, in the thermodynamic
limit, either the free energy of the original GREM with discrete
hierarchical structure, or of its continuous hierarchy analogue
(see (\ref{freeenfin}) below) analysed recently in much detail by
Bovier and Kurkova \cite{BoK}, and appearing also in earlier
studies of random heteropolymers\cite{PWW}.

The physical interpretation of our results also generalize the
discussion of the previous section, Sect. \ref{sect2end},  in a
natural and, we believe, rather beautiful way. Instead of one
localisation transition temperature $T_c$ where the particle
chooses a finite number of ``blobs'' of size $O(a)$ where the
potential is particularly deep, there appears $K$ different
transition temperatures, where the particle localizes on finer and
finer length-scales. The largest transition temperature $T_1$
corresponds to a condensation of the Boltzmann-Gibbs weight inside
a few blobs of large size $O(R)$, but the particle is still
completely delocalized {\it inside} each blob. As the temperature
is reduced, the REM condensation takes place over smaller blobs of
size $O(R^\nu)$ inside each already occupied large blobs, and this
scenario repeats itself as the temperature is reduced, each time
``zooming" in on a smaller scale \footnote{see \cite{microscope}
for a related discussion of the idea that temperature plays the
role of a microscope in the context of spin-glasses.}. To see this
most clearly we quote the simplest example beyond REM, the
two-scale logarithmic model characterized by the density of
exponents
$g^2(\nu)\rho(\nu)=g_2^2\delta(\nu)+g_1^2\delta(\nu-\nu_1)$, with
$0<\nu_1<1$. The system turns out to be described by two different
critical temperatures $T_1=\sqrt{g_1^2+g_2^2}>T_2=g_2$. As we will
be able to show, see Eq.(\ref{zoom}), in our model the probability
$\Pi(\nu)$ for two particles to be found at a distance ${\cal
D}=O(R^{2\nu})$ apart is given by:
\begin{equation}\label{zoom1}
\Pi(\nu)=
\left\{\begin{array}{l}\quad\quad\qquad\qquad\qquad\delta(\nu-1)\,,\,\,\,\quad\qquad\qquad\qquad\qquad\qquad
T>T_1\\ \quad\qquad
\left(1-\frac{T}{T_1}\right)\delta(\nu-\nu_1)+\frac{T}{T_1}\delta(\nu-1)\,
,\qquad\qquad\qquad
T_2<T\le T_1, \,\,\\
\left(1-\frac{T}{T_2}\right)\delta(\nu)+\left(\frac{T}{T_2}-\frac{T}{T_1}\right)
\delta(\nu-\nu_1)+\frac{T}{T_1}\delta(\nu-1)\, ,\quad 0\le T\le
T_2
\end{array}\right.
\end{equation}
The first two lines reproduce the former REM scenario, with $T_1$
standing for $T_c$ and the scale $a_1=O(R^{\nu_1})$ playing for
$T>T_2$ the role of the lowest discernible cutoff scale.  The last
line describes quantitatively the ``zooming in" from the scale
$a_1=O(R^{\nu_1})$ to the even smaller cutoff scale $a=O(R^0)$,
which becomes discernible below $T=T_2$ and dominates more and
more when $T\to 0$.

\subsection{Analytical results for $N \to \infty$}

We aim to compute the equilibrium free energy per degree of freedom of our model,
$F_{\infty}=\lim_{N\to\infty}F_N/N$, where $F_N$ is defined in Eq. (\ref{freeendef}).
The disorder average is performed in a standard way using the replica
trick. The replicated partition function $\langle Z^n\rangle_V\,$
is evaluated exactly for $1\le n\le N-1 $ in the large-$N$ limit
by the Laplace method, after exploiting a high symmetry of the
integrand stemming from the symmetry of the correlation function
Eq.(\ref{potential}). The replica limit $n\to 0$ is then performed
in the standard framework of the Parisi hierarchical ansatz. The
details of the corresponding analysis can be found in \cite{FS},
and we give below a summary of the most essential formulae.

From the point of view of the Schwarzian derivative criterion
recalled above, our models (\ref{scalingln}) and
(\ref{scalingln1}) are such that for any finite $R<\infty$, the
low-temperature phase is characterized by {\it continuous} FRSB
Parisi pattern with infinite level of hierarchy. This holds
invariably, even when we make the choice of a discrete set of
$K\ge 2$ distinct exponents $0<\nu_K<\ldots<\nu_1<1$. Indeed, the
function $S(u)$ defined in Eq.(\ref{2}) when calculated from
Eq.(\ref{scalingln1}) reads:
\begin{equation}\label{A}
S(u)=-3\int \int_{0}^{1}\frac{g^2(\nu) \rho(\nu)g^2(\nu') \rho(\nu')}{(u+s_{\nu}^2)^2(u+s_{\nu'}^2)^2}
\left[\frac{1}{(u+s_{\nu}^2)}-\frac{1}{(u+s_{\nu'})}\right]^2 \,d\nu \,d\nu',
\end{equation}
where we have used the short-hand notation
$s_{\nu}^2=a^2+R^{2\nu}$. The expression in the right-hand side of
(\ref{A}) is manifestly strictly negative, apart from the discrete
$K=1$ case when it is zero. Only in the thermodynamic limit $R\to
\infty$ shall we find that $S(u)\to 0$ for most values of $u$. In
this limit the system effectively recovers a GREM structure
corresponding to the replica symmetry breaking pattern with $K$
levels of hierarchy\footnote{An alternative construction of
multiscale logarithmic landscapes which show replica symmetry
breaking with exactly $K$ levels of Parisi hierarchy for {\it
finite} $R$ is proposed in the Appendix B. }.

For finite $R$, the low temperature phase is therefore
characterised by the existence of a nontrivial, non-decreasing
function $x(q),\, q\in [q_0,q_k]$, with the two parameters $q_0$
and $q_k$ satisfying the inequality $0\le q_0\le q_k\le q_d \equiv
R^2$. The corresponding $F_{\infty}$ can be written in terms of
only those two parameters, see the equation (58) of \cite{FS},
without explicit reference to $x(q)$. Here we find it more
convenient to introduce, along the line of the physical discussion
given above, two characteristic ``blob" sizes (actually size
squared) $d_{\min}=R^2-q_k,\,d_{\max}=R^2-q_0$ in terms of which:
\begin{eqnarray}\label{freeenglass}
\nonumber F_{\infty}&=&-\frac{T}{2}\ln{\left[2\pi e
d_{\min}\right]}+\frac{1}{2T}\left[f(d_{\min})-f(0)-d_{\min}f'(d_{\min})\right]
\\&+&\frac{f'(d_{\max})}{\sqrt{f''(d_{\max})}}-\int_{d_{\min}}^{d_{\max}}\sqrt{f''(u)}\,du,
\end{eqnarray}
where $d_{\min} \leq d_{\max}$ can be found
for a given temperature $T$ from the equations
\begin{equation}\label{minmax}
0 \leq d_{\min}=\frac{T}{\sqrt{f''(d_{\min})}},\quad
d_{\max}=R^2+\frac{f'(d_{\max})}{f''(d_{\max})} \leq R^2
\end{equation}
Finally, the Parisi order-parameter function, which takes the values between $0$ and $1$ and is the main measure of the
ultrametricity in the phase space, has the following shape
\begin{equation}\label{con4}
x(d)=-\frac{T}{2}\frac{f'''(d)}{[f''(d)]^{3/2}},\quad \forall
d\in[d_{\min},d_{\max}]\,\,.
\end{equation}
 where again we found convenient to perform the overall change $q\to d=R^2-q$ in comparison with
notations used in \cite{FS}. This function must be now {\it
non-increasing}, as follows from relating its derivative to the
probability $\pi(d)$ introduced earlier in Eq.(\ref{overlap}), see
the relation Eq.(\ref{overlap4}). Technically, this property is
precisely ensured by negativity of the Schwarzian derivative of
$f(d)$, see discussions around Eq.(\ref{2}).

The above solution is valid for the temperature range $0\le T\le T_{c}$, where the critical
(or de Almeida-Thouless) temperature
$T_{c}$ is given in terms of the largest blob size $d_{\max}$ as:
\begin{equation}\label{AT}
T_{c}=d_{\max}\sqrt{f''(d_{\max})}\,.
\end{equation}
Above this temperature the solution is replica-symmetric, corresponding to a
 delocalized phase for the particle: no particular region dominates the
partition function. The corresponding free energy is given by:
\begin{eqnarray}\label{freeensym}
\nonumber && F_{\infty}=-\frac{T}{2}\ln{\left[2\pi
d_{s}\right]}+\frac{1}{2T}\left[f(d_s)-f(0)\right]-\frac{T}{2}\frac{R^2}{d_s}
\end{eqnarray}
where $d_s$ satisfies
\begin{equation}\label{repsym}
d_{s}=R^2+\frac{d^2_{s}}{T^2}f'(d_{s})\, .
\end{equation}

We now consider specifically a correlation functions $f(u)$ of the
form (\ref{scalingln}). In what follows we will use the convenient
notations $z=({2\ln{R}})^{-1}$ and $y=z\ln{(u+a^2)}$. As noted
above, our multiscale logarithmic model ensures that  $\Phi'(y)\ge
0$ and $\Phi''(y)\ge 0$ for any $0<y<1$. We also will assume in
our analysis below that the function $\Phi''(y)$ is finite (
$0<\Phi''(y)<\infty$) and differentiable, but later on will relax
those conditions. Simple differentiation gives:
\begin{equation}\label{scal1}
f'(u)=-\frac{1}{u+a^2}\Phi'(y), \quad
f''(u)=\frac{1}{(u+a^2)^2}\left[\Phi'(y)-z\Phi''(y)\right],
\end{equation}
Our first goal is to find the largest blob size $d_{\max}$ from second equation in
Eq.(\ref{minmax}), and then to determine the critical
 temperature $T_{c}$. Introduce the scaling variable
$y_{\max}=z\ln{(d_{\max}+a^2)}$ and using (\ref{scal1}), we obtain
the following equation determining $y_{\max}$:
\begin{equation}\label{scal2}
e^{(1-y_{\max})/z}=\left(1-a^2e^{-y_{\max}/z}\right)
\left[1+\frac{\Phi'(y_{\max})}{\Phi'(y_{\max})-z\Phi''(y_{\max})}\right],
\end{equation}
Since we are interested in the thermodynamic limit $z\to 0$,
we can look for a solution $y_{\max}(z)$ as a power series of $z$.
One immediately checks that $y_{\max}(z)=1-z\ln{2}+O(z^2)$. This implies
that the largest blob size is of the order of the system radius:
$d_{\max} \approx {R^2}/{2}\gg a^2$ for $R\to \infty$.  Eq. (\ref{scal1}) and (\ref{AT}) then yield the
critical temperature given in the thermodynamic limit, which simply reads:
\begin{equation}\label{AT1}
T_{c}=\sqrt{\Phi'(1)}\, .
\end{equation}
Physically, at $T_c$, the space breaks up blobs of size $o(R)$ and only a
finite number of these blobs are visited by the particle. However, within each
blob, all sites are more or less equivalent. Now we can treat along the same
lines the first equation in (\ref{minmax}) to determine the smallest blob size $d_{\min}$ for $T < T_c$.
It can again be conveniently written in terms of the scaling variable $y_{\min}=z\ln{(d_{\min}+a^2)}$, such that:
\begin{equation}\label{scal3}
T=\left(1-a^2e^{y_{\min}/z}\right) \sqrt{\Phi'(y_{\min})-z\Phi''(y_{\min})}.
\end{equation}
This equation determines $y_{\min}\ge 0$ for any temperature $T<T_{c}$ and sample size $R=\exp{[1/2z]}$. In the
thermodynamic limit $z\to 0$, it is again natural to look for a
solution $y_{\min}$ as a power series of $z$, in which we only retain the first two terms:
$y_{\min}=\nu_*+cz+O(z^2)$. Assuming self-consistently that the solution
corresponds to $\nu_*>0$, we see that the first factor in
(\ref{scal3}) can be replaced with unity with exponential
accuracy. Due to our assumption on differentiability of the
function $\Phi'(y)$ we expand around $y=\nu_*$, and after a simple
calculation find $c=1$. This means that $d_{\min}$ behaves like
$d_{\min}=eR^{2\nu_*}$ for $R\to \infty$, where $\nu_*$ satisfies
the equation
\begin{equation}\label{main1}
T^2=\Phi'(\nu_*)\,.
\end{equation}
Since the function $\Phi'(y)$ is monotonously increasing for
$y>0$, we find that in the limit $R\to \infty $  (i.e. $z\to 0$),
the equation Eq.(\ref{main1}) must have a unique solution
$1>\nu_*(T)>0$ in the range of temperatures
$\sqrt{\Phi'(0)}=T_{\min} <T < T_{c}=\sqrt{\Phi'(1)}$. In this
regime, $d_{\min} \ll d_{\max}$. Physically, sites within blobs of
size $d_{\min}$ or smaller are not resolved by the particle, which
visits all of them more or less equally.

Now we can easily find the free energy $F_{\infty}$ in the
thermodynamic limit $z\to 0$. In particular, note that
Eq.(\ref{scal1}) implies that the last term in
Eq.(\ref{freeenglass}) can be conveniently written as:
\begin{equation}\label{integ}
I=-\int_{d_{\min}}^{d_{\max}}\left[(u+a^2)^2f''(u)\right]^{1/2}\frac{du}{u+a^2}=
-\frac{1}{z}\int_{y_{\min}}^{y_{\max}}\left[\Phi'(y)-z\Phi''(y)\right]^{1/2}\,dy
\end{equation}
In the temperature range $T_{\min}<T\le T_{c}$ we can substitute here
$y_{\min}=\nu_*(T)+z$ and $y_{\max}=1-z\ln{2}$, and expand in $z$ up to
linear terms. This gives:
\begin{equation}\label{integ2}
I=-2\ln{R}\int_{\nu_*}^{1}\sqrt{\Phi'(y)}\,dy+T_{c}(1+\ln2)
\end{equation}
In the same way we evaluate the remaining terms in
Eq.(\ref{freeenglass}), and finally find the leading and the
subleading terms for the equilibrium free energy:
${F_{\infty}}=\ln R \,{\cal F}+\delta F$, where the leading term
coefficient ${\cal F}$ is given by: \be \label{freeenfin} -{\cal
F}/{T}= \nu_*(T)+\frac{\left[\Phi(\nu_*)-\Phi(0)\right]}{T^2}
+\frac{2}{T}\int_{\nu_*}^1\sqrt{\Phi'(y)}\, dy, \quad
T_{\min}<T\le T_{c} \ee and the correction term is:
\be\label{freeenfin11} -\delta F/{T}=
\ln{\sqrt{2\pi}}+1-\frac{T_{c}}{T}\ln{2}-\frac{T_{\min}^2}{T^2}\ln{a},\quad
T_{\min}<T\le T_{c}. \ee For $T>T_{c}$ the solution of
(\ref{repsym}) in the limit $a\ll R\to \infty$ is given by:
\begin{equation}\label{repsym1}
d_s=R^2\frac{T^2}{T^2+T_{c}^2}\,,
\end{equation}
and substituting this to (\ref{freeensym}) we find that the free energy
components are given by:
\begin{eqnarray}\label{freeensyma}
&& -{\cal F}/{T}=1+\frac{\left[\Phi(1)-\Phi(0)\right]}{T^2} \,,\qquad \qquad T>T_c\,,\\
&& -\delta F/{T}=
\ln{\sqrt{2\pi}}+\frac{1}{2}\left(1+\frac{T^2_{c}}{T^2}\right)
\left[1-\ln{\left(1+\frac{T^2_{c}}{T^2}\right)}\right]-\frac{T_{\min}^2}{T^2}\ln{a}\,.
\end{eqnarray}

Finally, the analysis should be reconsidered for
$T<T_{\min}=\sqrt{\Phi'(0)}$, where $d_{\min} \sim a^2$ and the
particle localizes even on the smallest scales $O(a)$. Indeed,
assuming that generically $\Phi''(0) < \infty$ it is easy to see
from Eq.(\ref{main1}) that $\nu_*\approx (2T_{min}/\Phi''(0))
(T-T_{\min})\to 0$ as $T\to T_{\min}$. The solution
$d_{\min}=eR^{2\nu_*}$ is therefore invalid for $T<T_{\min}$ where
$\nu_*$ stays identically zero. Using Eqs. (\ref{minmax}) and
(\ref{scal1}) we find after a straightforward calculation the
following correct solution to Eq.(\ref{scal3}) as $z\to 0$ in this
range of temperatures:
\begin{equation}\label{mainREM}
d_{\min}=a^2 \frac{T}{T_{\min}-T}, \quad 0\le T<T_{\min},
\end{equation}
showing that $T_{\min}$ is indeed a delocalisation transition of
the REM type, above which the minimum blob size becomes much
larger than the small scale cut-off $a$. The leading term in the
free energy is given by $\nu_*\to 0$ limit of (\ref{freeenfin}):
\begin{equation} \label{freeREM}
{\cal F}=-2\int_{0}^1\sqrt{\Phi'(y)}\,dy, \quad 0\le T\le
T_{\min},
\end{equation}
Free energy corrections can also be computed, and for all
$T<T_{\min}$ are given by
\begin{equation} \label{freerem1}
-\delta
F/T=\ln{\sqrt{2\pi}}+\frac{T_{min}}{2T}-\frac{T_{c}}{T}\ln{2}+\left(1-2\frac{T_{min}}{T}\right)\ln{a}
+\frac{1}{2}\left(1+\ln{\frac{T}{T_{min}}}\right)
\end{equation}
\[
-\frac{1}{2}\left(1-\frac{T_{min}}{T}\right)^2\ln{\left(1-\frac{T}{T_{\min}}\right)}\,.
\]
Comparing Eqs. (\ref{freerem1}) and (\ref{freeenfin11}) we see
that the correction term is continuous at the delocalization
transition point $T=T_{min}$.

Last but not least, we can determine the thermodynamic limit of
the order-parameter function $x(d)$ given by Eq.(\ref{con4}),
which determines in a precise way how the particle localizes on
different scales. To leading order in $z$ we find
$f'''(u)=-2\Phi'(y)/u^3$ with $y=z\ln{(u+a^2)}$. Introduce again
the scaling variable $\nu=\frac{\ln{d}}{2\ln{R}}$ with
$d\in[eR^{2\nu_*},R^2/2]$. Denoting the Parisi order-parameter
function $x(d)$ expressed in terms of the new variable $\nu$ as
$X(\nu)$ we see that such function assumes the limiting form:
\begin{equation}\label{con4a}
X(\nu)=\frac{T}{\left[\Phi'\left(\nu\right)\right]^{1/2}},\quad
\forall \nu\in[\nu_*,1]\,\,.
\end{equation}

This completes our solution of the problem for the case of continuous function $\Phi(y)$.
At this point it is rather informative to consider the case of a discrete spectrum of exponents $\nu$, corresponding to
$K$ superimposed logarithmic potentials with:
\begin{equation}\label{disc}
g^2(\nu) \rho(\nu)=\sum_{i=1}^K\,g_i^2\,\delta(\nu-\nu_i),
\quad 0<\nu_K<\nu_{K-1}<\ldots<\nu_{1}<\nu_0=1\,,
\end{equation}
with $\delta(u)$ standing for the Dirac delta-functions. The corresponding $\Phi'(y)$ consists of steps:
$\Phi'(y)=\sum_{i=1}^Kg_i^2\theta(y-\nu_i)$. A simple
consideration shows that our earlier analysis for the values of
$d_{\max}$ and the critical temperature $T_{c}$ still
hold for such a case, so $d_{\max}=R^2/2$, and
$T_{c}=[\Phi'(1)]^{1/2}=\sqrt{g_1^2+g_2^2+\ldots g_K^2}$.
The equation (\ref{minmax}) used to determine $d_{\min}=R^{2y_{\min}}-a^2$ now takes the following form:
\begin{equation}\label{disc1}
T^2=\sum_{i=1}^Kg_{i}^2\frac{1-a^2e^{-y_{\min}/z}}{1+e^{(\nu_i-y_{\min})/z}},
\quad z=\frac{1}{2\ln{R}}.
\end{equation}
A little thought shows that the solution should always be in the
form $y_{\min}=\nu_p+c_pz$ for small $z$, where the index $p$ runs
successively through the values $1,...,K$ when decreasing
temperature from $T_{c}$ towards $T=0$. Introducing a decreasing
sequence of characteristic temperatures
$T_p=\sqrt{\sum_{i=p}^{K}g_i^2}$, we find the coefficients $c_p$
and the index $\nu_p$ for a given temperature:
\begin{equation}\label{disc2}
y_{\min}=\nu_p+z\ln{\frac{T^2-T_{p+1}^2}{T_{p}^2-T^2}}, \quad
T_{p+1}<T<T_p
\end{equation}
Thus, the value of $y_{\min}$ jumps (and thus the size of the
smallest frozen blobs $d_{\min}$)  when crossing each of the
temperatures $T_p$, $p=1,2,\ldots,K$ with the highest one being
$T_1=T_c$. Since $T_p$ and $\nu_p$ decrease as $p$ increases, it is
clear that the order-parameter function  $X(\nu)$ for a given
temperature $T < T_c$ is step-wise constant with jumps at each
$\nu_p$; the smaller $\nu$ (i.e. the smaller the size of the blobs),
the larger $X(\nu)$,  meaning that the condensation effect on the
scale $\nu$ is weaker and the Boltzmann weight becomes delocalized
for scales such that $X(\nu)\geq 1$.  Explicitly, in the temperature
range $ T_{p+1}<T \leq T_p$ we find \bea
X(\nu)&=&1+\left(\frac{T}{T_p}-1\right)
\theta(\nu-\nu_p)+\left(\frac{T}{T_{p-1}}-\frac{T}{T_p}\right)\theta(\nu-\nu_{p-1})+\ldots
\\
\ldots &+&
\left(\frac{T}{T_{1}}-\frac{T}{T_2}\right)\theta(\nu-\nu_1)-\frac{T}{T_1}\theta(\nu-1)\,.
\eea Invoking the relation Eq.(\ref{overlap4}) we then see that
the probability $\Pi(\nu)$ for two independent particles to end up
at an ultrametric separation $\nu$ is given for the present model
by \bea
\Pi(\nu)&=&\left(1-\frac{T}{T_p}\right)\delta(\nu-\nu_p)+\left(\frac{T}{T_{p}}-
\frac{T}{T_{p-1}}\right)\delta(\nu-\nu_{p-1})+\ldots
\\
\label{zoom}\ldots &+&
\left(\frac{T}{T_{2}}-\frac{T}{T_1}\right)\delta(\nu-\nu_1)+\frac{T}{T_1}\delta(\nu-1)\,,
\eea as long as $ T_{p+1}<T \leq T_p$, $p=1,2,\ldots K$. In
particular, in the simplest case of the two-scale model $K=2$
assuming $0=\nu_2<\nu_1<1$ for the exponents, we obtain the
probability distribution quoted in Eq.(\ref{zoom1}).

The expressions for $(y_{\min},y_{\max})$ suffice to calculate the
free energy expression in the thermodynamic limit. One easily
finds the leading order contribution to be $-F_{\infty}=\ln{R}\,
{\cal F}$, where in the phase with broken replica symmetry we have
for $p=1,2,\ldots,K-1$
\begin{equation}\label{freeendisc}
 -{\cal F}/T=\nu_p+\frac{2}{T}\sum_{i=1}^p(\nu_{i-1}-\nu_i)\,T_i+
\frac{1}{T^2}\sum_{i=p+1}^K(\nu_{i-1}-\nu_i)\,T^2_i, \quad
T_{p+1}<T<T_p,
\end{equation}
and finally for $p=K$
\begin{equation}\label{freeendisca}
 -{\cal F}/T=\nu_k+\frac{2}{T}\sum_{i=1}^K(\nu_{i-1}-\nu_i)\,T_i
, \quad 0<T<T_K=g_K\,.
\end{equation}
The corresponding replica symmetric expression valid for
$T>T_{1}=T_{c}$ is given by
\begin{equation}\label{freeendisc1}
-{\cal F}/T=1+\frac{1}{T^2}\sum_{i=1}^K(\nu_{i-1}-\nu_i)\,T^2_i\,.
\end{equation}
Interestingly, these expressions reproduce exactly, {\it mutatis
mutandis} the leading order free-energy expressions of Derrida's
GREM \cite{GREM}, with a particularly clear interpretation in terms
of particle localization inside smaller and smaller blobs as the
temperature is reduced. Corrections to the free energy can also be
found, but the corresponding expressions are rather cumbersome and
are not universal but model-dependent. In the Appendix B we provide
the explicit free energy expression for any value of $R$ for a
different model with $K-$step RSB, which has the same GREM-like
thermodynamic limit as the present model.

We end this section by a comment on the nature of that latter
model, which we believe deserves separate mentioning. Disordered
Hamiltonians usually analysed in spin-glass literature give rise
to either $K=1$ (1RSB), or to $K=\infty$ (FRSB) Parisi patterns,
see for example the results in the framework of the so-called
spherical model of spin glasses in \cite{sph,Talsph}. As is easy
to show, see \cite{CuLD} and also \cite{FS}, the general class of
the models of the type (\ref{potential})  includes the spherical
model as a special case. It is therefore can be of some
independent interest to provide an explicit example of a system of
this sort which has a $K-$ step version of the Parisi hierarchy as
an {\it exact} solution, for arbitrary $K\ge 1$ (see
\cite{Crisanti} for a recent example of a model with a $K=2$ RSB
solution). In the Appendix B we succeed in constructing such an
example in the framework of the model (\ref{potential}) of a
particle in a random potential when $N \to \infty$ even for {\it
finite values} of the sample radius $R<\infty$ (this case can be
looked at as corresponding to a {\it bona fide} p-(soft)spin model
with a spherical constraint). It comes as no surprise that in the
appropriately taken thermodynamic limit $R \to \infty$ this type
of models reproduce again the same GREM behaviour as elsewhere in
the present paper.

\subsection{Multifractality of the Boltzmann-Gibbs measure}

Important information about structure of the Gibbs-Boltzmann
equilibrium measure $p_{\beta}({\bf
r})=\frac{1}{Z(\beta)}\exp{-\beta V({\bf
r})},\,\,\beta=\frac{1}{T}$ can be extracted from the knowledge of
moments
\begin{equation}\label{BGmom}
\quad m_q=\int_{|{\bf r}|\le L} p^q_{\beta}({\bf r})\, d {\bf
r}=\frac{Z(\beta q)}{\left[Z(\beta)\right]^q}\,.
\end{equation}
In the thermodynamic limit of the sample volume $V_L \to \infty$
one expects typically
\begin{equation}\label{BGmom1}
\quad m_q\sim V_L^{-\tau_q}
\end{equation}
where the set of exponents $\tau_q$ reflects the spatial
organization of the Gibbs-Boltzmann weights. For example, if the
weights are of the same order of magnitude across the sample
volume, the normalization condition implies locally
$p_{\beta}({\bf r})\sim V_L^{-1}$, and a simple power counting
predicts the exponents $\tau_q=q-1$. In such a situation it is
conventional to speak about a {\it delocalised} measure. The
opposite case of a fully {\it localised} measure describes the
situation when essential Gibbs-Boltzmann weights concentrate in
the thermodynamic limit in a domain with the finite total volume
$V_{\xi}\ll V_L\to \infty$, and are vanishingly small outside that
domain. This situation is obviously characterized by trivial
exponents $\tau_{q>0}=0$ and $\tau_{q<0}=\infty$. Finally, in many
interesting situations the exponents $\tau_{q}$ may depend on $q$
nonlinearly, and in this case one commonly refers to the {\it
multifractality} of the measure. The Eqs.(\ref{BGmom}) and
(\ref{BGmom1})
 imply the following expression for the characteristic exponents $\tau_q$
 in the general case
\begin{equation}\label{BGmom2}
\quad \tau_q=|q|\beta{\cal F}(|q|\beta)-q\beta{\cal F}(\beta)
\end{equation}
relating them to the appropriately normalized free energy of the
system:
\begin{equation}\label{BGmom3}
{\cal F}(\beta)=-\lim_{V_L\to \infty}\frac{\ln{Z(\beta)}}{\beta
\ln{V_L}}\,.
\end{equation}

An alternative way of characterizing multifractality invokes the
so-called singularity spectrum function $f(\alpha)$. This function
characterizes the number  $dN(\alpha)=V_L^{f(\alpha)}d\alpha$ of
sites in the sample where the local Gibbs-Boltzmann measure scales
as $p_{\beta}({\bf r})\sim V_L^{-\alpha}$ in the thermodynamic
limit. The definition allows to extract the characteristic
exponents $\tau_q$ as
\begin{equation}\label{BGmom4}
\tau_q=-\lim_{V_L\to \infty}\frac{\ln{\int_{f(\alpha)\ge
0}e^{-\ln{V_L}[\alpha q-f(\alpha)]}\,d\alpha}}{\ln{V_L}}\,.
\end{equation}
Note, that the restriction of the integration range by the
condition $f(\alpha)\ge 0$ is necessary in order to remove the
rare events found in the vanishing number $dN(\alpha)\to 0$ of
sites in the thermodynamic limit\cite{ME}. This indeed ensures
that the extracted exponents $\tau_q$ characterize the typical
behaviour of the moments. Performing the $\alpha-$integration by
the Laplace method one finds that the two ways of characterizing
multifractality, by the set of exponents $\tau_q$ or by the
singularity spectrum $f(\alpha)$, turn out to be simply related by
a Legendre transform: $\tau_q=\alpha_* q-f(\alpha_*), \quad
q=f'(\alpha_*)$.

The expressions for the free energy calculated in the previous
section allow us to extract the multifractality exponents $\tau_q$
from Eq.(\ref{BGmom2}-\ref{BGmom3}) and to investigate the
corresponding singularity spectrum, see the relation
Eq.(\ref{BGmom4}). For the systems with a finite number $K$ of
levels of hierarchy the singularity spectrum $f(\alpha)$ turns out
to be piecewise-parabolic,  generalizing earlier results obtained in
the framework of REM-like model with a single scale logarithmic
correlations\cite{2d,CLD}. For the case of an infinite hierarchy of
scales in the landscape , i.e. for  a continuous set of exponents
$\nu$, the behaviour turns out to be rather rich and unusual. Below
we give a short account of the results for the example of continuous
function $\Phi'(y)$ assuming $\Phi'(0)=0$ for simplicity.

The associated singularity spectrum $f(\alpha)$ calculated via
Eq.(\ref{BGmom4}) is positive in an interval
$\alpha\in(\alpha_{min},\alpha_{max})$. The positions of the zeros
 $\alpha_{min},\alpha_{max}$  of the function $f(\alpha)$ are
given by
\begin{eqnarray}\label{zeroes}
\alpha_{min}=-\beta {\cal F}(\beta)-2\beta
\int_{0}^1\sqrt{\Phi'(y)}\,dy\,,\\ \alpha_{max}=-\beta {\cal
F}(\beta)+2\beta \int_{0}^1\sqrt{\Phi'(y)}\,dy\,.
\end{eqnarray}
The function $f(\alpha)$ is symmetric with respect to the midpoint
of the interval of interest,
$\alpha_m=(\alpha_{min}+\alpha_{max})/2=-\beta {\cal F}(\beta)>0$,
where it has the maximum $f(\alpha_{m})=1$ as expected. Close to
this maximum, namely, in the subinterval
$\alpha\in(\alpha_{-},\alpha_{+})$, where the endpoints
$\alpha_{\pm}=\alpha_m\pm 2A(\beta)\frac{T}{T_{c}}$ with
$A(\beta)=\beta^2 (\Phi(1)-\Phi(0))$ the singularity spectrum has
a simple parabolic shape:
\begin{equation}\label{parab}
f(\alpha)=1-\frac{1}{4A(\beta)}\,(\alpha-\alpha_m)^2,\quad
\alpha_-\le \alpha\le \alpha_{+}\,.
\end{equation}
In particular, at the boundaries $f(\alpha_{\pm})=1-\beta_{c}^2
\left(\Phi(1)-\Phi(0)\right)$. Note that in the REM-like limit
$\Phi(y)=g^2y$ we have $\alpha_{min/max}\to \alpha_{-/+}$ and the
parabolic behaviour extends to the whole interval of positivity of
$f(\alpha)$, in full agreement with the results of \cite{2d,CLD}.

At the same time for $\alpha\notin (\alpha_{-},\alpha_{+})$ the
GREM-like model may show a much richer multifractal structure
manifesting itself, in particular, via a quite unusual shape of
the singularity spectrum close to the zeros
$\alpha_{min},\alpha_{max}$, Eq.(\ref{zeroes}). To illustrate this
fact, we consider a broad class of functions $\Phi(y)$ behaving at
small arguments $y\ll 1$ as $\Phi(y)\approx C^2\,y^{2s+1}$, where
$s\ge 0$ and the coefficient $0<C<\infty$. In particular, in the
limiting case $s\to 0$ we are back to the old REM-like model. The
behaviour of $f(\alpha)$ in the vicinity of the endpoints
$\alpha_{min}$ or $ \alpha_{max}$ is dictated by the behaviour of
the multifractality exponents at large $q$, i.e. $\tau_{|q|\to
\infty}$. As is easy to see from Eq.(\ref{BGmom2}) this
asymptotics is in turn extracted from the knowledge of the
low-temperature behaviour of the free energy: $\beta{\cal
F}(\beta)|_{\beta\to \infty}$. To investigate that limit from
Eq.(\ref{freeenfin}) we need to know the low-temperature
asymptotics of the parameter $\nu_*$ related to $T$ via the
equation (\ref{main1}). In this way we find that for the chosen
$\Phi(y)$ the parameter $\nu_*(T)$ behaves as $\nu_*(T\to
0)\approx \left(\frac{T}{C\sqrt{2a+1}}\right)^{-1/s}$. After
simple algebra we find with the required accuracy
\begin{equation}\label{lowT}
-\beta{\cal F}(\beta)|_{\beta\to \infty}\approx 2\beta
\int_0^1\sqrt{\Phi'(y)}\,dy-\alpha_c(\beta)\,,
\end{equation}
where we introduced the short-hand notation
\begin{eqnarray}
\alpha_c(\beta)=\frac{2s^2}{(s+1)(2s+1)}(\beta
C\sqrt{2s+1})^{-\frac{1}{s}}\,.
\end{eqnarray}
Now the relation Eq.(\ref{BGmom2}) immediately yields the required
asymptotic behaviour of the multifractal exponents:
\begin{equation}\label{asymtau}
\tau_{q}=\left\{\begin{array}{c}
q\,\alpha_{min}-\alpha_c(\beta)\,q^{-\frac{1}{s}},\quad\mbox{for}\quad
q\to \infty\\
q\alpha_{max}-\alpha_c(\beta)|q|^{-\frac{1}{s}},\quad\mbox{for}\quad
q\to -\infty
\end{array}\right.
\end{equation}
where $\alpha_{min/max}$ are precisely the zeroes of $f(\alpha)$
given by Eq.\,(\ref{zeroes}). We thus see that the singular behaviour
revealed by (\ref{asymtau}) is very different from the
situation typical for other types of disordered
systems\cite{ME,MG} where one always observes a {\it precise}
linear behaviour $\tau_q=q\alpha_{min/max}$ starting from some
value of $|q|$ (see formula (2.42) in \cite{ME} and
discussions around it).  In particular, such anomalous asymptotics
is translated by the Legendre transform to the anomalous
singularity spectrum behaviour close to the left and right zero:
\begin{eqnarray}
f(\alpha)\approx
\frac{s+1}{s^{s/(s+1)}}\,\alpha_{c}(\beta)^{s/(s+1)}|\alpha-\alpha_{min/max}|^{\frac{1}{s+1}}\,,
\end{eqnarray}
We see that for any $s>0$ the derivative of the singularity spectrum
diverges when approaching zeroes as $f'(\alpha)\sim
|\alpha-\alpha_{min/max}|^{-\frac{s}{s+1}}\to \infty$. Note that
this is again very different from the standard behaviour observed in
other disordered systems where always $f'(\alpha)<\infty$ at zeros
of $f(\alpha)$ \cite{ME}.

\section{Discussion and Conclusion}
\label{sect4}

In this last section, we want to discuss several aspects of our model which
are certainly worth investigating further, among which: (i) the case of finite
dimensions; (ii) the dynamics in such a multiscale landscape and (iii) the relation
with multifractal random walks.

\subsection{Multiscale logarithmic landscape in finite dimensions}

As detailed at the end of Sect. \ref{sect2end}, the exact, $N \to \infty$ results for the single scale logarithmic potential
match perfectly with the results obtained by Carpentier and Le Doussal using RG, numerical and heuristic methods \cite{CLD} in
finite dimensions. There is no reason to doubt that these results are in fact exact in {\it all} dimensions $N \geq 1$. Although our model
is substantially more involved, we see that essentially the same physical mechanisms are at play in both models, in particular in the case
of a finite number $K$ of hierarchies.  Therefore, it is very tempting to conjecture that
the GREM behaviour revealed by above in the infinite-dimensional
setting should also hold in {\it all} spatial dimensions, down to $N=1$. It would be very interesting to see if the
corresponding RG and travelling wave formalism can be generalized to support this conclusion. In the case of finite $K$, this looks indeed quite feasible.

If this conjecture is  true, we would then have indeed explicitly constructed a Parisi landscape in finite dimensions fully in terms of {\it stationary} Gaussian processes.
How do we reconcile this with the ultrametric properties of the Parisi construction?  Consider the
following distance $D_R$ defined for any two points ${\bf r},{\bf r'}$ inside a sphere of the radius $R$ in the Euclidean
space of any dimension $N$:
\begin{equation}\label{distance}
D_R({\bf r},{\bf r'})=\frac{\ln{\left[|{\bf r}-{\bf r'}|^2+a^2\right]}}{2\ln{R}}, \quad 0< |{\bf r}|,|{\bf r'}|\le R
\end{equation}
Parameterizing $|{\bf r}| \equiv R^{\alpha({\bf r})}, \, 0\le \alpha \le 1$, we see that in fact
\begin{equation}\label{ultradistance}
\lim_{R\to \infty}D_R({\bf r},{\bf r'})=\max\{\alpha({\bf r}),\alpha({\bf r'})\}\,.
\end{equation}
One can easily check that the latter function used as a distance
converts the Euclidean sphere into a so-called ultrametric space:
every triangle will have at least two sides equal. We thus
conclude that our choice of the model corresponds to insisting
that the covariance of the random potential values in two points
in space should depend only on the ultrametric distance inside our
growing sphere. The original construct of the GREM by Derrida in
fact proceeds in a similar way, and therefore the coincidence
between our model of the GREM could have been, with hindsight,
anticipated. Indeed, Derrida started with $2^M$ random variables,
attached to the vertices of a hypercube of $2^M$ points supplied
with a tree structure and associated ultrametric distance
\cite{BoK}. The Gaussian random energies used for constructing the
Boltzmann weights were built from those ingredients in such a way
that their correlation function depended only on that ultrametric
distance. Although the methods used in this paper have very little
in common with techniques used by Derrida, or other authors
analysing GREM, the similarities between the two problems is
apparent. This line of reasoning suggests that the convergence to
the GREM limit could be indeed expected by invoking universality
arguments. A remarkable recent progress \cite{BoK} based on the
idea of Ruelle probability cascades \cite{Ruelle} has allowed one
to analyze Derrida's original GREM in full mathematical rigour. It
would be extremely interesting to see whether the Euclidean
version of the GREM analysed in the present paper is amenable to a
similar kind of rigorous analysis, without any reference to the
replica trick, powerful heuristically but still ill-defined
mathematically.

Another, more speculative aspect of the problem is also worth
mentioning here. In a recent work, Moore \cite{Moore}, argued that
low-temperature phases with 1RSB fail to survive in finite spatial
dimensions. This, in Moore's argument, is intimately related to a
generic absence of marginally stable modes in a 1RSB fluctuation
spectrum. To this end, the stability of the 1RSB low-temperature
phase in the present model (\ref{potential}) was investigated in
\cite{FS}, and found to be controlled by two eigenmodes, denoted
in \cite{FS} as $\Lambda_0^*$ and $\Lambda_K^*$ (see equations
(B.29) and (B.30) of the Appendix B of \cite{FS}). Generically
those two were demonstrated to be positive, but for the
logarithmic potential (\ref{2c}) both $\Lambda_0^*$ and
$\Lambda_K^*$ identically vanish everywhere in the low-temperature
phase. That property makes the associated 1RSB phase {\it
marginally stable}\footnote {This fact, though not explicitly
mentioned in \cite{FS}, immediately follows from definitions
(B.29) and (B.30) after substituting for $q_1-q_0=Q$ and
$q_d-q_1=y$ the expressions (74) and (79) of that paper.}. Thus,
the glass behaviour in the case of logarithmic correlations could
survive for finite $N$ due to marginality of the fluctuation
spectrum. This picture would be indeed in agreement with the
above-discussed RG results of \cite{CLD}. We believe that similar
marginality of fluctuations spectra should exist also in the
general $K-$step case of our model, and is related to the fact of
vanishing Schwarzian derivative, exactly or asymptotically in the
thermodynamic limit. The corresponding calculation looks
cumbersome, but is certainly feasible, see similar work in
\cite{CDT}, and is yet to be done. In general, any results in
these directions are highly desirable, with an ultimate goal of
performing perturbative expansions around $N=\infty$ limit which
remains to be one of outstanding challenging tasks.

\subsection{Diffusion in a multiscale logarithmic landscape}

The rich behaviour found in the thermodynamics of a single particle in a
random potential also has interesting dynamical counterparts. In the infinite dimension
limit, the problem of a Langevin particle in a random potential has been solved in details by
Franz and M\'ezard \cite{FM} and Cugliandolo and Le Doussal
\cite{CuLD}, for both the short-range and the long range cases. These results reveal
long-time relaxation, aging, and other effects typical of glassy dynamics.
The marginal (monoscale) logarithmic case was however not specifically studied in these papers.
For finite $N$, this marginal case was studied in \cite{BGLD,HK} using RG methods. The main
quantity of interest is the dynamical exponent $z$, defined as:
\be\label{diff}
\Delta^2(t)=\langle \left({\bf r}(t+t_0)-{\bf r}(t_0)\right)^2 \rangle \sim_{t \to \infty} t^{2/z}.
\ee
The result, conjectured to hold at all orders in perturbation theory and for any $N$, is that $z$ varies
continuously with the strength of the disorder $g$ and temperature $T$:
\be\label{z}
z = 2 + 2 \left(\frac{g}{T}\right)^2.
\ee
When $g=0$, one recovers the diffusion exponent $z=2$, as it should. It was noted by Castillo and Le Doussal \cite{CasLD} that this result cannot hold down to $T=0$ in $N=1$ dimension,
because it violates exact bounds. It was argued instead that below the static transition $T_c=g$ found in
Sect. \ref{sect2end}, the exponent in fact is modified and reads $z = 4T_c/T$. Therefore the static transition is also a dynamical transition in $N=1$.

The situation in higher dimensions is unclear, in particular it is not immediately clear that the dynamical transition where Eq. (\ref{z}) ceases to hold still coincides
with the static transition. This is another motivation for studying in details the limit $N \to \infty$. However,
it seems reasonable to conjecture that for the multiscale landscape model at high enough temperature, the
diffusion exponent becomes scale dependent. For a finite number $K$ of hierarchies, and for $T > T_1 \equiv T_c$, we expect to find:
\be
z(\Delta)=2 + 2 \left(\frac{T_p}{T}\right)^2,
\ee
where $T_p$ is the critical temperature associated to $\nu_p$ such that $R^{\nu_p} < \Delta < R^{\nu_{p-1}}$. At temperatures lower than $T_c$, it
would appear natural to conjecture that all levels such that $T < T_p$ follow the Castillo-Le Doussal scenario, whereas levels such that $T_p > T$ are still
ruled by the above exponent. However, the situation might be more complicated because the possibility of aging at low temperatures seems to have been overlooked
in \cite{CasLD}. In high enough dimensions, the analogy with the GREM suggests that the multiscale logarithmic model of the present paper might be a real space
realisation of the multi-level trap model introduced and studied in \cite{BDean}. We would then witness a very rich dynamical behaviour,
where all levels such that $T < T_p$ age (concerning large length scales), whereas small length scales, such that $T > T_p$, are still stationary \cite{BDean,microscope}. It would be extremely
interesting to study these aspects in more details.

\subsection{A Generalized Multifractal Random Walk}

The {\it mono}scale logarithmic landscape model in $N=1$ has in fact deep connections with the {\it multi}fractal Random Walk (MRW) construction of Bacry, Muzy \& Delour \cite{BMD,Wilmott}. If one treats the
coordinate $r$ as a time variable $t$, and call $X_t$ the position of a random walk at time $t$, the BMD model is defined by the following evolution equation:
\be
dX_t = m dt + \sigma_t dW_t \qquad \sigma_t=\sigma_0 \exp \beta V_t,
\ee
where $dW$ is the usual Brownian process, $m$ a drift and $V_t$ is the logarithmically correlated Gaussian process considered above (possibly shifted such that $\langle e^{2\beta V_t} \rangle_V = 1$).
It was shown by BMD that the resulting process is multifractal in the time regime $a \ll t \ll R$. For $m=0$, all odd moments vanish and even moments are given by \cite{BMD}:
\be\label{multif}
\langle (X_t - X_{t+\tau})^n \rangle = M_n \tau^{\zeta_n} \quad {\mbox{with}} \quad \zeta_n=\frac{n}{2}-(\beta g)^2 \frac{n(n-2)}{2},
\ee
as long as $n <(\beta g)^{-2}$ beyond which all moments diverge: the distribution of increments $\Delta=(X_t - X_{t+\tau})$ has a power-law tail $\Delta^{-1-\mu}$ with an exponent given by:
\be
\mu = \left(\frac{T}{T_c}\right)^2.
\ee
The connection with the thermodynamical problem addressed above is clear: the MRW is a standard random walk subordinated
to a stochastic time $s_t$, such that:
\be
s_{t+\tau}-s_t = \int_t^{t+\tau} du \, e^{2\beta V_u},
\ee
which is a restricted partition function at temperature $T/2$.

The generalisation to a multiscale logarithmic process is very
natural. Taking $V_t = \sum_{\nu=1}^K V_{\nu,t}$ as in Eq.
(\ref{multidef}) above, one constructs a Generalized MRW with, in
the limit $R \to \infty$, $K$ different {\it epochs} such that
$\nu_p < \ln \tau/\ln R < \nu_{p-1}$ such that the moments of the
walk are given by the above multifractal scaling, Eq.
(\ref{multif}), but with a different multifractal spectrum: \be
\label{mulfrspec} \zeta_n^{(p)}=\frac{n}{2}-
\left(\frac{T_p}{T}\right)^2 \frac{n(n-2)}{2}. \ee This formula
can be given a meaning in the continuous limit as well. In this
case the exponents $\zeta_n^{(p)}$  acquire a continuous
dependence on $\ln \tau$: they are still given by
Eq.(\ref{mulfrspec}) but with the ratio $T/T_p$ replaced with the
Parisi function $X(\nu)$, $\nu=\ln{\tau}/\ln R$. It would be
interesting to understand in more details the extreme value
statistics of the GMRW, following the method introduced in
\cite{CLD,Kozh}.

\subsection{Final remarks}

The model we have introduced appears to contain a very rich
phenomenology, and deserves in our opinion further investigations.
Many points would require a more rigorous mathematical
investigation, concerning in particular the finite dimensional
version of the model and its dynamical properties, in particular
in the continuous hierarchy case $K \to \infty$. We believe that
our construction sheds an important light on the understanding of
replica symmetry breaking: clearly the fact that Parisi's
ultrametric scheme is relevant for a translationally invariant,
Gaussian potential in finite (even one) Euclidean dimension is
very satisfying, if only for pedagogical reasons. A better
intuition on the Parisi solution for the SK model for a large but
finite number of spins $M$, and the recently reported scaling $K
\sim M^{1/6}$ \cite{Moore2}, might also be within reach.

\subsection*{Acknowledgements}
This paper is dedicated to David Sherrington on the occasion of his
65th birthday. The model studied here is of course in close
filiation with the ideas and methods David contributed to develop
and popularize.

An essential part of this research was performed during the first
author's stay at the Institute of Theoretical Physics, Cologne
University, Germany. YF appreciates kind hospitality extended to him
during that period, as well as the financial support from the
Alexander von Humboldt foundation through Bessel Research Award. The
research in Nottingham was supported by grant EP/C515056/1 from
EPSRC (UK). This project was initiated during the workshop on Random
Matrix Theory held in Jagellonian University, Cracow, May 2007. We
thank the organisers for this opportunity. We also want to thank
Mike Moore for interesting discussions.

\section*{Appendix A: Calculation of the overlap function, Eq.(\ref{overlap})}

In the framework of the replica trick we represent the
normalization factor in the product of the Boltzmann-Gibbs weights
as $1/Z^2=\lim_{n\to 0}Z^{n-2}$. This trick allows the disorder
average to be performed explicitly,  and (\ref{overlap}) takes the
form \bea  \pi({\cal D})=\lim_{n\to 0} & \int_{|{\bf r}_a|\le
L,\forall a}&\left\langle\,e^{-{\beta}\sum_{a=1}^n\,V({\bf
r}_a)}\right\rangle_{V}
 \delta\left({\cal D}-\frac{1}{2N}[{\bf r}_1-{\bf
 r}_2]^2\right)\,\prod_{a=1}^n\,d{\bf r}_a
\\ \nonumber
=\lim_{n\to 0}
 e^{\frac{\beta^2}{2}Nnf(0)}&\int_{|{\bf r}_a|\le R\sqrt{N}}&e^{N\beta^2\sum_{a<b}
f\left(\frac{1}{2N}({\bf x}_{a}-{\bf x}_b)^2\right)}
\delta\left({\cal D}-\frac{1}{2N}[{\bf r}_1-{\bf
 r}_2]^2\right)
 \,\prod_{a=1}^n\,d{\bf r}_a
\eea

At the next step we exploit the $O(N)$ rotational invariance of
the integrand and, assuming $N>n$, introduce the scalar products
${\bf r}_a {\bf r}_b=q_{ab}$ as new integration variables, see
\cite{FS} for a general description of the method. After further
rescaling $q_{ab}\to Nq_{ab}$ the integral in the right-hand side
of the equation takes the form
\begin{equation}\label{overlap3}
{\cal C}_{N,n} N^{Nn/2}\int_{D_Q}
\left(\mbox{det}Q\right)^{-(n+1)/2} e^{- N\beta\Phi_n (Q) }
\delta\left({\cal D}-\frac{1}{2}(q_{11}+q_{22})+q_{12}\right)\, dQ
\end{equation}
where
\begin{equation}\label{repham1}
 \Phi_n (Q)=-
 \frac{1}{2\beta}\ln{(\det{Q})}-\beta\sum_{a<b}
f\left[\frac{1}{2}(q_{aa}+q_{bb})-q_{ab}\right]
\end{equation}
and the integration domain $D_Q$ over the matrix $Q$ with entries
$q_{ab}$ is already $N-$independent: $D_Q=\{Q\ge 0,\, q_{aa}\le
\,R^2,\, a=1,\ldots n\}$. The proportionality constant ${\cal
C}_{N,n}$ is also known.

The form of the integrand in Eq.(\ref{overlap3}) is precisely the one
required for the possibility of evaluating the replicated
partition function in the limit $N\to \infty$ by the Laplace
method. Taking into account the
permutational symmetry of the integrand with respect to the
replica indices, we find after a straightforward calculation
\begin{equation}
\pi({\cal D})=\lim_{n\to 0}\frac{1}{n(n-1)}\sum_{a\ne
b}\delta\left({\cal
D}-\frac{1}{2}(q_{aa}+q_{bb})+q_{ab}\right)|_{Q\in \mbox{\small
stationary point}}\,.
\end{equation}
To perform the replica limit explicitly we rely upon the
assumption of validity of the hierarchical ansatz for the
stationary-point solution suggested by Parisi, see a description
in the context of the present model in Appendix A of \cite{FS}.
Denoting $m_l$ the size of the blocks in the Parisi scheme, we
find in the standard way (cf. Eq.(46) of \cite{FS}):
\begin{equation}
\pi({\cal D})=\lim_{n\to 0}\sum_{l=0}^k(m_{l+1}-m_{l})\delta({\cal
D}- q_d+q_l)=\int_0^{q_d}\delta({\cal D}-
q_d+q)x'(q)\,dq=x'(q_d-{\cal D})\,,
\end{equation}
where $x(q)$ is a non-trivial non-decreasing Parisi function
characterizing the low-temperature phase. In the present model it
turns out that the parameter $q_d$ is related to the effective
radius of the sample as $q_d=R^2$. It is
actually more convenient to use the variable $d=R^2-q$ rather than
$q$ itself, as we do in the main body of the paper. Accordingly,
we pass from $x(q)$ to the non-increasing function $x(d)$ to
which $\pi({\cal D})$ is especially simply related:
\begin{equation}\label{overlap4}
\pi({\cal D})=-\frac{d}{d{\cal D}}x({\cal D})\,,
\end{equation}
which is the counter-part of the standard interpretation of $x(q)$ in terms
of the overlap probability in spin-glasses (\cite{MPV}).

The above construction is valid for any choice of random Gaussian
potential with any covariance function of the form
Eq.(\ref{potential}). For the case of multiscale logarithmic
potential studied in the present paper the argument $d$ of the
function $x(d)$ generically covers the interval $d\in
[d_{min},d_{max}]$, with $d_{min}\sim R^{2\nu_*}$ and $d_{max}\sim
R^2$, where the value $\nu_*$ determined for a given temperature
$T$ by Eq.(\ref{main1}). Accordingly, it is natural to pass from
the Parisi function $x(d)$ to its counterpart $X(\nu),\,\, \nu\in
[\nu_*,1] $ by replacing $d\to R^{2\nu}$. As is clear, $\nu$ is
precisely the "ultrametric" separation between the two positions
related to the scaled squared Euclidean distance ${\cal D}$ as
$\nu=\frac{1}{2\ln{R}}\ln{{\cal D}}|_{R\to \infty}$, cf.
Eq.(\ref{ultradistance}). It is then immediate to check that the
probability $\Pi(\nu)$ for two independent particles to end up at
an ultrametric separation $\nu$ is related to the function
$X(\nu)$ in precisely the same way as $\pi({\cal D})$ to $x(d)$:
\begin{equation}\label{overlap5}
\Pi(\nu)=-2\ln{R}\,\frac{d}{d\ln{(\cal D})}x({\cal D})|_{{\cal
D}=R^{2\nu}}=-\frac{d}{d\nu} X(\nu) \,.
\end{equation}

\section*{Appendix B: A model for K-step replica symmetry breaking}

The idea of the construction comes from an observation that the
most general function $f(u)$ with locally vanishing Schwarzian
derivative $S(u)=0$, see Eq.(\ref{2}), is given by
$f(u)=f-Au-g^2\ln{(u+a^2)}$. Apparently, linear in $u$ term
violates the requirement $f'(u\to \infty)\to 0$ and for this
reason was discarded by us earlier. Now we can try, by choosing
constants $f$ and $A$ judiciously, to construct a covariance
function $f(u)$ globally from the above local patches in such a
way, that the composite solution will have the vanishing $S(u)$
everywhere, except in a finite number of points. As the derivative
$x'(d)$ of the Parisi order-parameter function from
Eq.(\ref{con4}) is proportional to the Schwarzian derivative
$S(d)$, this construction should provide us with a
piecewise-constant order-parameter function required for the
Parisi pattern with $K$ levels of hierarchy. For the sake of
simplicity we discuss below in detail the simplest non-trivial
case $K=2$. The generalization to arbitrary $K\ge 2$ will be
apparent.

We find it more convenient to work with the structure function
$\phi(u)=f(0)-f(u)$ rather than with $f(u)$ itself. The structure
function obviously vanishes at the origin, providing an additional
condition $\phi(0)=0$ used to specify all the constants in our
construction uniquely.

 Consider the model of a particle
in a Gaussian random potential with the structure function taken
in the form
\begin{equation}\label{K2a}
-\phi(u)=\left\{\begin{array}{c} f_1-A_1u-t_1^2\ln{(u+a_1^2)},
\quad u\ge u_*\\f_2-A_2u-t_2^2\ln{(u+a_2^2)}, \quad 0\le u\le u_*
\end{array}\right.\,,
\end{equation}
where the positive parameters $t_1,t_2,a_1,a_2$ are considered to
be given, and chosen satisfying the inequalities:
\begin{equation}\label{ineq}
t_1>t_2\quad \mbox{and}\quad \frac{a_1^2}{t_1}>\frac{a_2^2}{t_2}.
\end{equation}
In contrast, the value of the "breakpoint" $u_*$, as well of the
constants $f_1,f_2,A_1,A_2$ are unknown and should be specified in
terms of $a_{1,2}$ and $t_{1,2}$. Note that for $\phi(u)$ to have
the meaning of a structure function requires $A_{1,2}$ to be
non-negative, see eq. (\ref{longranged}).

 As we are to use such a function for building the Parisi
order-parameter $x(d)$ according to Eq.(\ref{con4}), and such
$x(d)$ can take only finite values between $0$ and $1$,  one has
to ensure that $\phi(u)$ is continuous together with at least two
derivatives everywhere, including the breakpoint $u_*$. The
requirement of  continuity of $\phi''(u)$ at $u_*$ immediately
fixes the value of the breakpoint:
\begin{equation}\label{break}
u_*=\frac{\frac{a_1^2}{t_1}-\frac{a_2^2}{t_2}}{\frac{1}{t_2}-\frac{1}{t_1}}.
\end{equation}
The consistency of the procedure requires $u_*>0$ which is
precisely the case due to the imposed conditions Eq.(\ref{ineq}).

In the same way the continuity of $\phi'(u)$ at $u_*$ given by
(\ref{break}) allows us to relate $A_1$ to $A_2$ via
\begin{equation}\label{bb}
A_2=A_1+\frac{(t_1-t_2)^2}{a_1^2-a_2^2}
\end{equation}
 At the
next step the continuity of $\phi(u)$ at $u_*$ relates $f_1$ to
$f_2$ as
\begin{equation}\label{bb2}
 f_1=f_2+
\frac{a_2^2t_1-a_1^2t_2}{a_1^2-a_2^2}(t_1-t_2)+t_1^2\ln{t_1}-t_2^2\ln{t_2}
+(t^2_1-t^2_2)\ln{\frac{a_1^2-a_2^2}{(t_1-t_2)}}\,.
\end{equation}
The constant $f_2$ can be found from the condition $\phi(0)=0$,
which fixes $f_2=t_2^2\ln{a_2^2}$. Finally one should have in mind
that although for a finite sample $x<R^2$,  to have a well-defined
model for any $x<\infty$ we should impose the condition
$\phi'(u\to \infty)=0$ as elsewhere in the paper. This necessarily
implies $A_1=0$, and fixes all the constants of the model
uniquely. Note that $A_2$ is indeed positive in view of
Eq.(\ref{bb}) and the inequality $a_1^2>a_2^2$.

Having specified the structure function $\phi(u)$, hence the
covariance $f(u)$, we now can repeat the procedure of finding
$d_{\min},d_{\max}$, and hence the free energy of the model.
Assuming $d_{\max}>u_*$ we can solve the second of equations
Eq.(\ref{minmax}) and find $d_{\max}=\frac{R^2-a_1^2}{2}$. Such
solution is indeed the valid one for sufficiently large sample
sizes satisfying
$R^2>R^2_{m}=[a_1^2(t_1+t_2)-2a_2^2t_1]/(t_1-t_2)$. Assuming the
condition is satisfied, the transition to the phase with broken
replica symmetry occurs at the temperature found from
Eq.(\ref{AT}) which is given by
\begin{equation}\label{AT3}
T_c=t_1\frac{R^2-a_1^2}{R^2+a_1^2}\,.
\end{equation}
Then the solution to the first of equations Eq.(\ref{minmax})
reads:
\begin{equation}\label{qmin11}
d_{\min}=\left\{\begin{array}{c}a_2^2\,\frac{T}{t_2}/{\left(1-\frac{T}{t_2}\right)},
\quad d_{\min}<u_*,\\
a_1^2\frac{T}{t_1}/\left(1-\frac{T}{t_1}\right)\,, \quad
u_*<d_{\min}
\end{array}\right.\,,
\end{equation}
and further requiring consistency of this solution with the
expression Eq.(\ref{break}) finally yields:
\begin{equation}\label{qmin22}
d_{\min}=\left\{\begin{array}{c}a_2^2\,\frac{T}{t_2}/\left(1-\frac{T}{t_2}\right),
\quad T\le T_{\min},\\
a_1^2\,\frac{T}{t_1}/\left(1-\frac{T}{t_1}\right)\,, \quad
T_{\min}\le T<T_c
\end{array}\right.\,,
\end{equation}
where the "breakpoint temperature" $T_{\min}$ is given by
\begin{equation}\label{AT4}
T_{\min}=t_1\,t_2\,\,\frac{a_1^2/t_1-a_2^2/t_2}{a_1^2-a_2^2}\,.
\end{equation}
It is easy to see that indeed $t_2<T_{\min}<T_c<t_1$ for $R>R_m$,
so such a solution is consistent. Note also that $d_{\min}(T)\le
u_*$ for $T<T_{\min}$, and $d_{min}(T_c)=d_{max}$ as expected.

A simple calculation shows that the Parisi order-parameter
function $x(d)$ in the temperature range $0\le T<T_{\min}$ indeed
consists of two perfect steps:
\begin{equation}\label{qmin33}
x(d)=\left\{\begin{array}{c}\frac{T}{t_2} \quad\mbox{for}\quad d_{\min}\le d<u_*,\\
\frac{T}{t_1}\, \quad\mbox{for}\quad u_*<d<d_{\max}
\end{array}\right.\,.
\end{equation}
whereas in the range $T_{\min}<T<T_c$ the value $d_{\min}$ exceeds
the breakpoint $u_*$, and hence only a single step survives:
\begin{equation}\label{qmin44}
x(q)= \frac{T}{t_1}\, \quad\mbox{for}\quad d_{\min}<d<d_{\max}\,.
\end{equation}

Now we can easily calculate the free energy for $T<T_c$  from
Eq.(\ref{freeenglass}). In particular, as we have for $0\le
T<T_{\min}$
\begin{equation}\label{qmin55}
\sqrt{f''(u)}=\left\{\begin{array}{c}\frac{t_2}{u+a_2^2}\,, \quad d_{\min}\le u<u_*,\\
\frac{t_1}{u+a_1^2}\,, \quad u_*<u<d_{\max}
\end{array}\right.\,
\end{equation}
we find that in this range of temperatures
\[
I=-\int_{d_{\min}}^{d_{\max}}\left[f''(u)\right]^{1/2}du=-t_2\ln{\left(\frac{u_*+a_2^2}{d_{\min}+a_2^2}\right)}
-t_1\ln{\left(\frac{d_{\max}+a_1^2}{u_*+a_1^2}\right)}\,.
\]
At the same time, for $T_{\min}<T\le T_c$ we have
$\sqrt{f''(u)}=\frac{t_1}{u+a_1^2}$ so that
\[
I=-t_1\ln{\left(\frac{d_{\max}+a_1^2}{d_{\min}+a_1^2}\right)}\,.
\]
The free energy Eq.(\ref{freeenglass}) is then given by
\begin{eqnarray}\label{qmin66}
&&-\frac{F_{\infty}}{T}=\ln{\sqrt{2\pi
e}}-\frac{f_1}{2T^2}+\frac{1}{2}\ln{\frac{T}{t_1}}+\frac{t_1}{T}\left[\frac{1}{2}+\ln{\frac{R^2+a_1^2}{2}}\right]
\\ \nonumber  &+&
\frac{1}{2}\left(1+\frac{t_1^2}{T^2}-\frac{2t_1}{T}\right)\ln{\frac{a_1^2}{1-\frac{T}{t_1}}},
\quad T_{\min}\le T \le T_c
\end{eqnarray}
and
\begin{eqnarray}\label{qmin77}
\nonumber  && -\frac{F_{\infty}}{T}=\ln{\sqrt{2\pi
e}}-\frac{f_2}{2T^2}+\frac{1}{2}\ln{\frac{T}{t_2}}-\frac{t_2}{T}\left[\frac{1}{2}
-\ln{(u_*+a_2^2)}\right]
+\frac{t_1}{T}\left[\frac{1}{2}-\ln{(u_*+a_1^2)}\right]
\\  && +\frac{t_1}{T}\left[\frac{1}{2}+\ln{\frac{R^2+a_1^2}{2}}\right]+
\frac{1}{2}\left(1+\frac{t_2^2}{T^2}-\frac{2t_2}{T}\right)\ln{\frac{a_2^2}{1-\frac{T}{t_2}}},
\quad 0\le T \le T_{\min}
\end{eqnarray}
One can check that at the breakpoint temperature $T=T_{\min}$ the free
energy is continuous.

Finally, for $T>T_c$ the replica-symmetric expression for $d_s$
obtained from Eq.(\ref{repsym}) is given by
\begin{equation}
d_s=\frac{1}{2(1+t_1^2/T^2)}\left[R^2-a_1^2+\sqrt{(R^2+a_1^2)^2+4R^2a_1^2t_1^2/T^2}\right]
\end{equation}
and the free energy is obtained by substituting this to
Eq.(\ref{freeensym}), and remembering $d_s>u_*$. One can check
that $d_s|_{T\to T_c}=(R^2-a_1^2)/2=d_{\max}$, and the free
energy is again continuous at the transition temperature $T_c$.

All the above expressions were derived for finite sample sizes,
provided $R>R_m$. Let us now investigate for this model the
thermodynamic limit $R\to \infty$, taken at a fixed temperature
$T$. As elsewhere in the paper, we assume the scaling
$a_{1}=R^{\nu_1},a_{2}=R^{\nu_2}$, with $0<\nu_2<\nu_1<1$, which
implies $a_1\gg a_2$ in the thermodynamic limit. We immediately
find that the critical temperature tends to $T_c=t_1$, and
$d_{\max}\to R^2/2, \, d_s\to R^2/(1+T^2_{c}/T^2)$. Similarly, the
breakpoint temperature $T_{\min}\to t_2$, and the breakpoint
itself is given by the limiting value
$u_*=R^{2\nu_1}/(T_c/T_{\min}-1)$. We also need to know that
$f_2=2\nu_2T_{\min}^2\ln{R}$ and
$f_1=2\left[\nu_1(T_c^2-T_{\min}^2)+\nu_2T_{\min}^2\right]\ln{R}$.
The leading contributions to the free energy found from
Eqs.(\ref{qmin66}) and (\ref{qmin77}) are given by
\begin{equation}\label{qmin88}
-\frac{F_{\infty}}{T\ln{R}}=\left\{\begin{array}{c}\nu_2+\frac{2T_{\min}}{T}(\nu_1-\nu_2)+\frac{2T_c}{T}(1-\nu_1)
,\quad 0\le T\le T_{\min}\\ \nu_1+
\frac{2T_c}{T}(1-\nu_1)+(\nu_1-\nu_2)\frac{{T^2_{\min}}}{T^2},
\quad T_{\min}\le T\le T_c\\
1+(1-\nu_1)\frac{T_c^2}{T^2}+(\nu_1-\nu_2)\frac{T_{\min}^2}{T^2},
\quad T\ge T_c\end{array}\right.
\end{equation}
These expressions are identical to $K=2$ case of GREM
free energy (\ref{freeendisc}) and (\ref{freeendisc1}), upon
identification $T_{\min}\equiv T_2, T_c\equiv T_1$, showing that the
two models are indeed in the same universality class in the
thermodynamic limit. The corrections are however already
model-dependent.

 We further note that the parameter
$d_{\min}$ given by Eq.(\ref{qmin11}) develops in the limit $R\to
\infty$ characteristic divergencies at the breakdown temperature
$T_{\min}$ as well as at $T_c$ (cf. (\ref{disc2})):
\begin{equation}\label{qmin11a}
d_{\min}=\left\{\begin{array}{c}\frac{T}{T_{\min}}\,R^{2\nu_2}/\left(1-\frac{T}{T_{\min}}\right),
\quad 0\le T<T_{\min},\\ \frac{T}{T_c}\,
R^{2\nu_1}/\left(1-\frac{T}{T_c}\right)\,, \quad T_{\min}<T<T_c
\end{array}\right.\,,
\end{equation}
The corrections to the free energy however do not have any
divergencies as the logarithmic terms come with vanishing
prefactors.

The $K=2$ construction discussed above has obvious generalization
to any $K\ge 2$. We choose the structure function in the form
\begin{equation}\label{K2aa}
-\phi(u)=f_p-A_p\,u-t_p^2\ln{(u+a_p^2)}, \quad u_*^{(p)}\le u\le
u^{(p-1)}_*,\,\, p=1,2,\ldots, K
\end{equation}
where the positive parameters $t_1,\ldots,t_K; a_1,\ldots,a_K$ are
given, and chosen satisfying the inequalities:
\begin{equation}\label{ineqa}
t_1>t_2>\ldots>t_k\quad \mbox{and}\quad
\frac{a_1^2}{t_1}>\frac{a_2^2}{t_2}>\ldots>\frac{a_K^2}{t_K}.
\end{equation}
Those inequalities will ensure that a strictly decreasing sequence
of the breakpoints $u_*^{(0)}=R^2> u^{(1)}_*> u^{(2)}_*>\ldots
>u^{(K-1)}_*>0$ can be found from the conditions of continuity of
$\phi''(u)$. Requiring also continuity of $\phi'(u)$ and $\phi(u)$
relates $A_p$ to $A_{p-1}$ and $f_p$ to $f_{p-1}$ for any $p$.
Finally we set $\phi(0)=0$ which gives $f_K=t_K^2\ln{a_K^2}$, and
put $A_1=0$ to satisfy the behaviour at $u\to \infty$. This choice
fixes all the parameters of the model uniquely, and ensures
$A_p>0$. Considering the effective radius $R$ exceeding some
minimal value $R_m\sim a_1^2$, the model will have the sequence of
"break temperatures" $T_c=T_1>T_2>\ldots>T_K>0$, with the
order-parameter function consisting of $l$ decreasing steps for
$T_{l-1}<T<T_l$, with the values at those steps given by
$x(d)=T/t_i,\,\mbox{for}\,\, u^{(i)}_*<d<u^{(i-1)}_*
\,i=1,2,\ldots,l$, where we make a convention $u^{(0)}_*\equiv
d_{\max},\,d^{(l)}_*\equiv d_{\min}$. Taking the thermodynamic
limit $R\to \infty$ with the the powerlaw scaling of the cutoffs
as $a_i=R^{2\nu_i}$ reproduces exactly the GREM equations
(\ref{freeendisc}) and (\ref{freeendisc1}).

\end{document}